\documentclass[aps,prl,twocolumn,groupedaddress,nofootinbib]{revtex4-2}
\usepackage{mathtools}
\usepackage{tikz}
\usepackage{stmaryrd}
\usetikzlibrary{arrows}
\usepackage{bbm}
\usetikzlibrary{decorations.markings}

\usepackage{amsmath}
\usepackage{amssymb}
\usepackage{xspace}
\usepackage{xcolor}

\newcommand{\lra}{\leftrightarrow}

\newcommand{\fchar}[1]{\hat{\bf #1}}

\newcommand{\circS}{\Delta}
\newcommand{\dcircS}{\tilde{\Delta}}
\newcommand{\circf}[3]{\circS^{#2}_{#1#3}}
\newcommand{\circVVV}{\circf{v}{v}{v}}
\newcommand{\circVBB}{\circf{v}{b}{b}}
\newcommand{\circBVB}{\circf{b}{v}{b}}
\newcommand{\circBBV}{\circf{b}{b}{v}}
\newcommand{\circBBB}{\circf{b}{b}{b}}
\newcommand{\circBVV}{\circf{b}{v}{v}}
\newcommand{\circVBV}{\circf{v}{b}{v}}
\newcommand{\circVVB}{\circf{v}{v}{b}}
\newcommand{\circFGH}{\circf{f}{g}{h}}

\newcommand{\circABC}{\circf{\alpha}{\beta}{\gamma}}

\newcommand{\circZPPM}{\circf{z^+}{z^+}{z^-}}
\newcommand{\circZPMP}{\circf{z^+}{z^-}{z^+}}
\newcommand{\circZPMM}{\circf{z^+}{z^-}{z^-}}
\newcommand{\circZMPP}{\circf{z^-}{z^+}{z^+}}
\newcommand{\circZMPM}{\circf{z^-}{z^+}{z^-}}
\newcommand{\circZMMP}{\circf{z^-}{z^-}{z^+}}

\newcommand{\circZpmZmpZpm}{\circf{z^{\pm}}{z^{\mp}}{z^{\pm}}}
\newcommand{\circZpmZmpZmp}{\circf{z^{\pm}}{z^{\mp}}{z^{\mp}}}
\newcommand{\circZpmZpmZmp}{\circf{z^{\pm}}{z^{\pm}}{z^{\mp}}}

\newcommand{\circZaZbZc}{\circf{z^{\alpha}}{z^{\beta}}{z^{\gamma}}}

\newcommand{\dcircZk}{\dcircS_\vk}
\newcommand{\dcircZp}{\dcircS_\vp}
\newcommand{\dcircZq}{\dcircS_\vq}

\newcommand{\vb}{{\bf{b}}}
\newcommand{\vf}{{\bf{f}}}
\newcommand{\vg}{{\bf{g}}}
\newcommand{\vh}{{\bf{h}}}
\newcommand{\vk}{{\bf{k}}}
\newcommand{\vm}{{\bf{m}}}
\newcommand{\vmA}{{\bf{m}_1}}
\newcommand{\vmB}{{\bf{m}_2}}
\newcommand{\vmC}{{\bf{m}_3}}

\newcommand{\vp}{{\bf{p}}}
\newcommand{\vq}{{\bf{q}}}

\newcommand{\vv}{{\bf{v}}}

\newcommand{\vzpm}{{\bf{z}}^\pm}
\newcommand{\vzmp}{{\bf{z}}^\mp}

\newcommand{\fb}{\hat{\bf b}}

\newcommand{\ff}{\hat{\bf f}}
\newcommand{\fg}{\hat{\bf g}}
\newcommand{\fh}{\hat{\bf h}}

\newcommand{\fv}{\hat{\bf {v}}}

\newcommand{\fz}{\hat{\bf z}}
\newcommand{\fzp}{\hat{\bf z}^+}
\newcommand{\fzm}{\hat{\bf z}^-}
\newcommand{\fzpm}{\hat{\bf z}^\pm}
\newcommand{\fzmp}{\hat{\bf z}^\mp}

\newcommand{\ftsf}{{\mathcal{T}}}

\newcommand{\ftsfCarg}[3]{#1\!\overset{#2}{\to}\!#3}
\newcommand{\ftsfform}[3]{\ftsf^{#2}_{#1#3}}
\newcommand{\ftsfC}[6]{\ftsfform{#1}{#2}{#3}(\ftsfCarg{#4}{#5}{#6})}

\newcommand{\kpqz}{\vk+\vp+\vq=0}

\newcommand{\pat}{\partial_t}

\newcommand{\reqref}[1]{Eq. \eqref{#1}}
\newcommand{\reqrefs}[2]{Eqs. \eqref{#1} and \eqref{#2}}
\newcommand{\reqrefss}[2]{Eqs. \eqref{#1}-\eqref{#2}}
\newcommand{\reqrefssne}[2]{\eqref{#1}-\eqref{#2}}

\newcommand{\kinE}{{\mathcal{E}}^v}
\newcommand{\magE}{{\mathcal{E}}^b}
\newcommand{\totE}{{\mathcal{E}}^T}
\newcommand{\elsE}{{\mathcal{E}}^{z^\pm}}
\newcommand{\elsEpm}{\elsE}
\newcommand{\elsEp}{{\mathcal{E}}^{z^+}}
\newcommand{\elsEm}{{\mathcal{E}}^{z^-}}

\newcommand{\somEf}{{\mathcal{E}}^f}

\newcommand{\somEg}{{\mathcal{E}}^g}
\newcommand{\somEh}{{\mathcal{E}}^h}

\newcommand{\beq}{\begin{equation}}
\newcommand{\eeq}{\end{equation}}

\newcommand{\beqa}{\begin{eqnarray}}
\newcommand{\eeqa}{\end{eqnarray}}

\newcommand{\fig}[1]{Fig. \ref{#1}}

\renewcommand{\Im}{\mathrm{Im}}
\renewcommand{\Re}{\mathrm{Re}}

\newcommand{\tsfIMterm}[6]{(#6\cdot#2_#5)(#1_#4\cdot#3_#6)}

\begin{document}

%\preprint{}

%\title{The mode-to-mode energetic transfers in triply periodic incompressible magnetohydrodynamics are unique}
%\title{\textcolor{red}{From triad interactions to pair exchanges in incompressible magnetohydrodynamics}}
\title{Pair-exchange dynamics of magnetohydrodynamic turbulence}


\author{Jean-Mathieu Teissier}
\email[]{jm.teissier@astro.physik.tu-berlin.de}
%\thanks{}
\affiliation{Technische Universität Berlin, ER 3-2, Hardenbergstr. 36a, D-10623 Berlin, Germany}

\author{Wolf-Christian M\"uller}
\affiliation{Technische Universität Berlin, ER 3-2, Hardenbergstr. 36a, D-10623 Berlin, Germany}
\affiliation{Max-Planck/Princeton Center for Plasma Physics, United States of America}


\date{\today}

\begin{abstract}
Nonlinear interactions in incompressible magnetohydrodynamic (MHD) and Navier-Stokes turbulence involve three spatial fluctuations conservatively exchanging, e.g., energy. The mathematical structure of a triad interaction does not allow to unambigously determine the specific
contribution of each of the participating fluctuations to the resulting exchanges. Presently, only additional assumptions allow for the deconstruction
of triad interactions into a sum of pair exchanges. This work proves that in incompressible MHD the simultaneous nonlinear transfer of cross-helicity
sufficiently constrains the triad dynamics such that a mathematically unique pair-exchange representation of nonlinear energy and cross-helicity transfer
is possible.

\end{abstract}

%\keywords{}


\maketitle

The intricate nonlinear dynamics of incompressible turbulence has been posing serious obstacles to the development of a complete and self-consistent
theory of this deterministically chaotic flow state (e.g. \cite{leslie:book,orszag:stattheory,monin_yaglom:book2} with \cite{kraichnan:dia,kraichnan:lhdia} as an important partial exception).
A full theoretical understanding of the conservative exchange of ideal invariants, e.g. energies or helicities, between turbulent fluctuations
is a less ambitious undertaking that still meets enormous difficulties.

Most importantly, in the discrete set of Fourier modes, which represents a finite, triply periodic volume of turbulent fields, this transfer process is caused by quadratic nonlinearities in the underlying partial differential equations. In such a case, most prominently found in the
Navier-Stokes equations governing the incompressive motion of electrically neutral, Newtonian fluids, such an exchange always involves three
interacting Fourier modes. To date, efforts to reduce this process
to a sum of mutual pair interactions remain on the level of approximations, e.g. \cite{alexakis_mininni_pouquet:eflowdiag}, or lack stricter mathematical
justification \cite{verma:review}. A related work \cite{PSV19} will be discussed below.
A pair-exchange representation would be an important advance, since it allows to treat turbulent dynamics with well-established
theoretical and modelling approaches such as rate equation networks \cite{beck_mueller.networkmodel}.

This Letter presents a relatively simple mathematical framework that renders it possible to prove the existence and uniqueness of nonlinear pair-exchanges
in the case of incompressible magnetohydrodynamics (MHD). This extension of the Navier-Stokes fluid model corresponds to a single-fluid approximation
of an electrically conductive medium, e.g. a plasma, subject to electric current densities and magnetic fields. It is valid on spatiotemporal
scales much larger than those characteristic for the constituent plasma particles. The present work exploits additional constraints imposed by
the presence of the magnetic field, {in particular, the conservative transfer of cross helicity}. 

The dimensionless equations of incompressible ideal MHD read:

\beqa
\label{eq:mhdS}\pat \vv &=&-\vv\cdot \nabla \vv + \vb \cdot \nabla \vb -\nabla p^*,\\
\label{eq:mhdb}\pat \vb &=&-\vv \cdot \nabla \vb + \vb \cdot \nabla \vv,\\
\label{eq:mhdE}\nabla \cdot \vv &=&\nabla \cdot \vb=0,
\eeqa

with $\vv$ the fluid velocity, $\vb$ the magnetic field and 
$p^*=p+b^2/2$ the total pressure, including the thermal pressure, $p$.

\newcommand{\vkfs}{\vq}

Considering a triply periodic MHD system and writing $\fv_\vkfs, \fb_\vkfs$ the Fourier coefficients of the velocity and magnetic fields at wave vector $\vkfs$, the kinetic and magnetic energetic contents $\kinE_\vq=\frac12 |\fv_\vq|^2, \magE_\vq=\frac12 |\fb_\vq|^2$  evolve as:

\def\hypera{\tikz[baseline=-2.]{
\draw (0,0.25) -- (0.25,0.);
\draw (0,-0.25) -- (0.25,0.);
\draw[->] (0.25,0.) -- (0.5,0.);}%
}

\def\smallhypera{\tikz[baseline=-2.]{
\draw (0,0.15) -- (0.15,0.);
\draw (0,-0.15) -- (0.15,0.);
\draw[->] (0.15,0.) -- (0.3,0.);}%
}

%https://tex.stackexchange.com/questions/582875/typeset-a-fraction-without-the-fraction-bar
\newcommand{\nobarfrac}{\genfrac{}{}{0pt}{}}
\newcommand{\fhyplett}{\mathcal S}
\newcommand{\fhypraw}[3]{{\fhyplett}^{\prescript{#2}{#1}{\scriptscriptstyle #3}}}
\newcommand{\fhyparg}[6]{\Big(\nobarfrac{\fchar{#2}_#5}{\fchar{#1}_#4} \hypera \fchar{#3}_#6\Big)}
\newcommand{\fwhyparg}[3]{\Big(\nobarfrac{#2}{#1} \hypera #3\Big)}
\newcommand{\fhyptsfC}[6]{\fhypraw{#1}{#2}{#3}\fhyparg{#1}{#2}{#3}{#4}{#5}{#6}}

\newcommand{\basefunc}{{\mathcal D}}
\newcommand{\fhypbudget}[3]{\fhyplett^{\nobarfrac{#2}{#1} \smallhypera #3}}

\newcommand{\fhypvgv}[3]{{\fhyplett}_{\kinE}^{-\vv\cdot\nabla \vv}\fwhyparg{#1}{#2}{#3}}
\newcommand{\fhypbgb}[3]{{\fhyplett}_{\kinE}^{\vb\cdot\nabla \vb}\fwhyparg{#1}{#2}{#3}}
\newcommand{\fhypvgb}[3]{{\fhyplett}_{\magE}^{-\vv\cdot\nabla \vb}\fwhyparg{#1}{#2}{#3}}
\newcommand{\fhypbgv}[3]{{\fhyplett}_{\magE}^{\vb\cdot\nabla \vv}\fwhyparg{#1}{#2}{#3}}

\newcommand{\crossH}{{\cal H}^C}
\newcommand{\wtowHc}[3]{{{\cal T}_{\crossH}}(\ftsfCarg{#1}{#2}{#3})}
\newcommand{\wtowEt}[3]{{{\cal T}_{\totE}}(\ftsfCarg{#1}{#2}{#3})}
\newcommand{\wtowEzp}[3]{{{\cal T}_{\elsEp}}(\ftsfCarg{#1}{#2}{#3})}
\newcommand{\wtowEzm}[3]{{{\cal T}_{\elsEm}}(\ftsfCarg{#1}{#2}{#3})}

\newcommand{\fhypHcvgv}[3]{{\fhyplett}_{\crossH}^{-\vv\cdot\nabla \vv}\fwhyparg{#1}{#2}{#3}}
\newcommand{\fhypHcbgb}[3]{{\fhyplett}_{\crossH}^{\vb\cdot\nabla \vb}\fwhyparg{#1}{#2}{#3}}
\newcommand{\fhypHcvgb}[3]{{\fhyplett}_{\crossH}^{-\vv\cdot\nabla \vb}\fwhyparg{#1}{#2}{#3}}
\newcommand{\fhypHcbgv}[3]{{\fhyplett}_{\crossH}^{\vb\cdot\nabla \vv}\fwhyparg{#1}{#2}{#3}}

\newcommand{\fhypggg}{{\fhyplett}_{\somEh}^{\pm\vg\cdot\nabla \vf}\fwhyparg{\vmA}{\vmB}{\vmC}}

\newcommand{\seesupmat}{see supplementary material}
\newcommand{\supmat}{supplementary material}
\newcommand{\supeqss}[2]{supplementary material, \reqrefss{#1}{#2}}

\beqa
\label{eq:pattrikin} \pat \kinE_\vq&=&\frac12\sum_{\kpqz} \Big(\fhypvgv{\vk}{\vp}{\vq}+\fhypbgb{\vk}{\vp}{\vq}\Big),\\
\label{eq:pattrimag} \pat \magE_\vq&=&\frac12\sum_{\kpqz} \Big(\fhypvgb{\vk}{\vp}{\vq}+\fhypbgv{\vk}{\vp}{\vq}\Big),
\eeqa

where the sums run over all wave vectors $\vp, \vq$ forming a triangle with $\vk$.
For an incompressible flow, the pressure does not contribute to energetic exchanges and thus does not appear in \reqrefs{eq:pattrikin}{eq:pattrimag}.
In this illustrative example, the transfer functions
$\mathcal{S}$ express the joint energetic effect of interactions involving modes at wave vectors \vk\  and \vp\  with a mode at \vq.
They are given by 
\beqa
\label{eq:hypervgv} \fhypvgv{\vk}{\vp}{\vq}&=&-\Im(\tsfIMterm{\fv}{\fv}{\fv}{\vk}{\vp}{\vq}+\tsfIMterm{\fv}{\fv}{\fv}{\vp}{\vk}{\vq}),\\
\fhypbgb{\vk}{\vp}{\vq}&=&\Im(\tsfIMterm{\fb}{\fb}{\fv}{\vk}{\vp}{\vq}+\tsfIMterm{\fb}{\fb}{\fv}{\vp}{\vk}{\vq}),\\
\label{eq:hyperfuncC} \fhypvgb{\vk}{\vp}{\vq}&=&-\Im(\tsfIMterm{\fb}{\fv}{\fb}{\vk}{\vp}{\vq}+\tsfIMterm{\fb}{\fv}{\fb}{\vp}{\vk}{\vq}),\\
\label{eq:hyperbgv} \fhypbgv{\vk}{\vp}{\vq}&=&\Im(\tsfIMterm{\fv}{\fb}{\fb}{\vk}{\vp}{\vq}+\tsfIMterm{\fv}{\fb}{\fb}{\vp}{\vk}{\vq}).
\eeqa

\newcommand{\genfhypbudget}{\fhypbudget{\ff_\vk}{\fg_\vp}{\fh_\vq}}
The elementary contributions to the transfer functions in this representation can be written more generally as
$\basefunc(\ff_\vk,\fg_\vp,\fh_\vq,\vq)=\Im(\tsfIMterm{\ff}{\fg}{\fh}{\vk}{\vp}{\vq})$.

In summary, the energetic exchanges are governed by triad interactions: three modes at different wave vectors forming a triangle interact with one another. In a triangle, $\kpqz$, there are exactly four triad MHD interactions governing the kinetic and magnetic energy transfers: one involving three velocity modes $\{\fv_\vk,\fv_\vp,\fv_\vq\}$ and three combining one velocity mode and two magnetic modes $\{\fv_\vk,\fb_\vp,\fb_\vq\}, \{\fb_\vk,\fv_\vp,\fb_\vq\}$ and $\{\fb_\vk,\fb_\vp,\fv_\vq\}$.

Neglecting the magnetic field in Eqs. \eqref{eq:mhdS}-\eqref{eq:mhdE} (Navier-Stokes case) and writing 
$\fhypbudget{\ff_\vk}{\fg_\vp}{\fh_\vq}$ for the rate of change of energy of type $h$ at $\vq$ through the triadic interaction between $\{\ff_\vk,\fg_\vp,\fh_\vq\}$,
detailed conservation of kinetic energy in each individual triad interaction 
is the well-known consequence, e.g. \cite{kraichnan:dia}, of 

\beq
\label{eq:detailedvvv}\fhypbudget{\fv_\vk}{\fv_\vp}{\fv_\vq}+\fhypbudget{\fv_\vp}{\fv_\vq}{\fv_\vk}+\fhypbudget{\fv_\vq}{\fv_\vk}{\fv_\vp}=0.
\eeq

In the MHD case, detailed conservation of kinetic energy is generally lost as it can be exchanged with magnetic energy by the  additional 
nonlinearities.
However,
the triad interaction of velocity modes still redistributes kinetic energy conservatively.
Similarly, for MHD triads with two magnetic modes and a velocity one, e.g. $\{\fb_\vk,\fb_\vp,\fv_\vq\}$ (\supeqss{eq:supmat_vkvpvq}{eq:supmat_vqbkbp}):

\beq
\label{eq:detailedvbb}\fhypbudget{\fb_\vk}{\fb_\vp}{\fv_\vq}+\fhypbudget{\fb_\vp}{\fv_\vq}{\fb_\vk}+\fhypbudget{\fv_\vq}{\fb_\vk}{\fb_\vp}=0.
\eeq

Thus, detailed conservation of the sum of kinetic and magnetic energy in MHD is a consequence of \reqrefss{eq:detailedvvv}{eq:detailedvbb}. 

Since each triad interaction conserves the total energy individually, the following analysis focuses on one triangle of wave vectors, $\kpqz$, and thus four possible triads for the energetic transfers.

The formulation of nonlinear transfers in terms of $\genfhypbudget$ does not allow for
a straightforward reduction to individual pair exchanges with unambigously defined sources and targets in terms of wave vectors or energy reservoirs. 

\newcommand{\clrt}{white}
\newcommand{\rnode}{0.3}
\newcommand{\ltri}{3.25}
\newcommand{\htri}{0.866*\ltri}

\newcommand{\xk}{0}
\newcommand{\yk}{0}
\newcommand{\xp}{0.5*\ltri}
\newcommand{\yp}{\htri}
\newcommand{\xq}{\ltri}
\newcommand{\yq}{0}

\newcommand{\xc}{0.5*\ltri}
\newcommand{\yc}{0.289*\ltri}

\newcommand{\mart}{0.15*\ltri}

\newcommand{\rrnode}{0.7}
\newcommand{\rvnode}{0.3}
\newcommand{\rEnode}{0.325}
\newcommand{\rbnode}{0.3}
\newcommand{\spnode}{0.35}

\newcommand{\xvk}{\xk-0.5*\spnode}
\newcommand{\yvk}{\yk+0.866*\spnode}
\newcommand{\xvp}{\xp+\spnode}
\newcommand{\yvp}{\yp}
\newcommand{\xvq}{\xq-0.5*\spnode}
\newcommand{\yvq}{\yq-0.866*\spnode}

\newcommand{\xbk}{\xk+0.5*\spnode}
\newcommand{\ybk}{\yk-0.866*\spnode}
\newcommand{\xbp}{\xp-\spnode}
\newcommand{\ybp}{\yp}
\newcommand{\xbq}{\xq+0.5*\spnode}
\newcommand{\ybq}{\yq+0.866*\spnode}

\newcommand{\ec}{-0.3*\ltri}

\newcommand{\xcvvv}{\xc}
\newcommand{\ycvvv}{\yc}

\newcommand{\xcvbb}{\xc+0.940*\ec}
\newcommand{\ycvbb}{\yc+0.643*\ec}
\newcommand{\xcbbv}{\xc-0.766*\ec}
\newcommand{\ycbbv}{\yc+0.342*\ec}
\newcommand{\xcbvb}{\xc+0.173*\ec}
\newcommand{\ycbvb}{\yc-0.984*\ec}

\begin{figure}
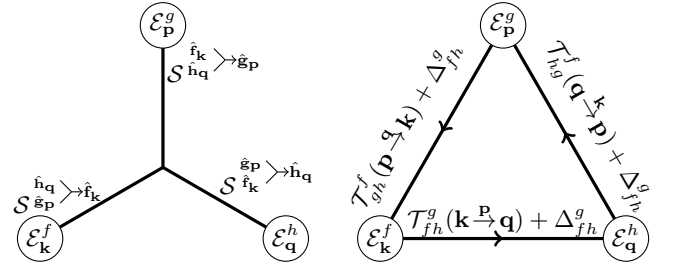
 
\newcommand{\sxi}{4.5}
\begin{tikzpicture} 
\begin{scope}[shift={(0,0)}]
\input{tikz_hyper_fgh.tex}
\end{scope}
\begin{scope}[shift={(\sxi,0)}]
\input{tikz_graph_fgh.tex}
\end{scope}
\end{tikzpicture} 
 \caption{\label{fig:hgg} {General triad interaction (left) and equivalent pair exchange view with an \textit{a priori} arbitrary circulating term (right).}}
 \end{figure}

A possible pair-exchange approach, which can be regarded an asymptotic limit of transfer approximations based on sharp Fourier-space partitioning
\cite{alexakis_mininni_pouquet:eflowdiag}, employs mode-to-mode transfer functions \cite{DVE01,PSV19}. This interaction model assumes that only two of the three modes in a triad,
e.g. $\ff_\vk$ and $\fh_\vq$, exchange energy.
Therefore, the third mode does not give rise to an additional effective transfer channel within
the triad, although it is analytically coupled to the pair-transfer rate.
Following established naming conventions, it is called \textit{mediator}.

The symbol $\ftsfC{f}{g}{h}{\vk}{\vp}{\vq}$ for a pair-transfer function expresses the signed rate of transfer of energy from type $f$ at $\vk$ to type $h$ at $\vq$. The direction of the transfer, i.e. the role of source and target is encoded in the sign of $\mathcal{T}$.

\newcommand{\Rbudget}{R1}
\newcommand{\Ranti}{R2}

\newcommand{\pattriint}[3]{\pat^{#1\overset{#2}{\to}#3}}

The pair-transfer functions must fulfill two basic requirements: first, the resulting transfers must be consistent with the
governing equations,
\beq
\tag{\Rbudget}\label{req1}
\genfhypbudget=\ftsfC{f}{g}{h}{\vk}{\vp}{\vq}+\ftsfC{g}{f}{h}{\vp}{\vk}{\vq}. 
\eeq

Second, the energy transfer rate between two Fourier modes with a fixed mediator changes sign if the direction of the transfer is reversed:
\beq
\tag{\Ranti}\label{req2}
\ftsfC{h}{g}{f}{\vq}{\vp}{\vk}=-\ftsfC{f}{g}{h}{\vk}{\vp}{\vq}.
\eeq

A physically meaningful decomposition of $\genfhypbudget$ in a sum of {pair exchanges}, see \fig{fig:hgg}, requires mathematical uniqueness.
However, the determination of  pair transfer functions $\ftsf$ from the knowledge of the $\fhyplett$-functions corresponds to solving a linear system (cf. \cite{PSV19}):

\newcommand{\matA}{\underline{\boldsymbol A}}

\beq
\label{eq:sysA}
\underbrace{
\begin{pmatrix}
1 & 0 & -1 \\
-1 & 1 & 0 \\
0 & -1 & 1
\end{pmatrix}}_{\matA}
\begin{pmatrix}
\ftsfC{f}{g}{h}{\vk}{\vp}{\vq} \\
\ftsfC{g}{h}{f}{\vp}{\vq}{\vk} \\
\ftsfC{h}{f}{g}{\vq}{\vk}{\vp}
\end{pmatrix}
=
\begin{pmatrix}
\fhypbudget{\ff_\vk}{\fg_\vp}{\fh_\vq}\\
\fhypbudget{\fg_\vp}{\fh_\vq}{\ff_\vk}\\
\fhypbudget{\fh_\vq}{\ff_\vk}{\fg_\vp}
\end{pmatrix},
\eeq

with a singular matrix $\matA$ of rank 2.
Since this is compatible with the right-hand side for incompressible MHD (cf. \reqrefss{eq:detailedvvv}{eq:detailedvbb}), there exists, \textit{without imposing additional constraints}, an infinite number of possible pair transfer functions.

All transfer solutions differ from a particular solution by a virtual or ``circulating'' transfer between the three modes without impact
on the net change of energy of any mode in the triad. 
Thus, for incompressible MHD, a particular solution can be expressed with the help of the $\basefunc$-function, giving the
general result (cf. \cite{PSV19}):

\beqa
\label{eq:generalvv} \ftsfC{v}{v}{v}{\vk}{\vp}{\vq}&=&-\basefunc(\fv_\vk,\fv_\vp,\fv_\vq,\vq)+\circVVV(\fv_\vk,\fv_\vp,\fv_\vq),\\%\vk,\vq),\\
\label{eq:generalbv} \ftsfC{b}{b}{v}{\vk}{\vp}{\vq}&=&\basefunc(\fb_\vk,\fb_\vp,\fv_\vq,\vq)+\circBBV(\fb_\vk,\fb_\vp,\fv_\vq),\\%\vk,\vq),\\
\label{eq:generalvb} \ftsfC{v}{b}{b}{\vk}{\vp}{\vq}&=&\basefunc(\fv_\vk,\fb_\vp,\fb_\vq,\vq)+\circVBB(\fv_\vk,\fb_\vp,\fb_\vq),\\%\vk,\vq),\\
\label{eq:generalbb} \ftsfC{b}{v}{b}{\vk}{\vp}{\vq}&=&-\basefunc(\fb_\vk,\fv_\vp,\fb_\vq,\vq)+\circBVB(\fb_\vk,\fv_\vp,\fb_\vq),%\vk,\vq),
\eeqa

with the $\circABC$ denoting the `circulating' transfers (see \fig{fig:hgg}). They are formally also functions of the wavevectors $\vk,\vp,\vq$. This dependency is dropped here to simplify the notation.

For a given triad, which is completely determined by the participating modes and their wavevectors, there is exactly one
circulating transfer term $\circABC$ defined as reference via the modes $\alpha$, $\beta$ and $\gamma$ at
wavevectors $\vk$, $\vp$ and $\vq$, respectively.

The requirements \eqref{req1} and \eqref{req2} for pair exchanges have consequences for the circulating transfer functions.
Due to \eqref{req1}, $\circABC$ is invariant under cyclic permutation of the three modes. Due to \eqref{req2}, swapping
the position of any of those modes reverses the sign of $\circABC$. 

An additional key requirement, which is not a consequence of \eqref{req1} or \eqref{req2} (\supeqss{eq:RTa}{eq:RTb}) but which is essential for the proof below, comes
from the following assertion: since $\fhypbudget{\ff_\vk}{\fg_\vp}{\fh_\vq}=0$ whenever a mode in the triad is zero,
it appears reasonable that the three pair exchanges vanish in this case as well (as implicitely assumed in \cite{PSV19}). This yields:

\beq
\label{eq:R3}\tag{R3}
\circABC(\ff_\vk,\fg_\vp,\fh_\vq)=0\text{ if any amplitude vanishes.}
\eeq

The work \cite{PSV19} remarks that, under the additional physical assumption that the advective $-\vv \cdot \nabla \vb$ and the stretching term $\vb\cdot\nabla \vv$ in the induction equation are solely responsible for, respectively, magnetic$\lra$magnetic and kinetic$\lra$magnetic energetic transfers in Fourier space as well, the transfers involving the magnetic field become uniquely defined, with $\circVBB=\circBVB=\circBBV=0$.

The present work proves that this additional assumption is not necessary. Indeed, it is a consequence of \reqref{eq:R3}. Furthermore, the last remaining circulating transfer function for kinetic$\lra$kinetic exchanges, $\circVVV$, is identically zero, as well, i.e., all energetic pair exchanges do exist and are unique in incompressible MHD.

The proof requires consideration of the cross helicity transfers, as well as the formulation of the MHD equations in terms of Els\"asser variables. Relations between the circulating transfer functions in both equivalent views give mathematical constraints that can only be fulfilled if these are identically zero.

The cross helicity $\crossH=\int \vv\cdot\vb$ is an ideal quadratic invariant of the MHD equations. Four triads, complementary to those for the energetic transfers, govern its transfer: either three magnetic modes $\{\fb_\vk,\fb_\vp,\fb_\vq\}$ or two velocity and a magnetic one: $\{\fb_\vk,\fv_\vp,\fv_\vq\}, \{\fv_\vk,\fb_\vp,\fv_\vq\}$ and $\{\fv_\vk,\fv_\vp,\fb_\vq\}$. Thus, in all generality, pair transfers of cross helicity are given by (\supeqss{eq:patcrossH}{eq:vvbwtot}):

\beq
\label{eq:crossHvb} \wtowHc{\vk}{\vp}{\vq}=\Big(\ftsfform{b}{b}{b}+\ftsfform{b}{v}{v}+\ftsfform{v}{b}{v}+\ftsfform{v}{v}{b}\Big)(\ftsfCarg{\vk}{\vp}{\vq}),
\eeq

with:
\beqa
\ftsfC{b}{b}{b}{\vk}{\vp}{\vq}&=&\basefunc(\fb_\vk,\fb_\vp,\fb_\vq,\vq)+\circBBB(\fb_\vk,\fb_\vp,\fb_\vq),\\
\ftsfC{b}{v}{v}{\vk}{\vp}{\vq}&=&-\basefunc(\fb_\vk,\fv_\vp,\fv_\vq,\vq)+\circBVV(\fb_\vk,\fv_\vp,\fv_\vq),\\
\ftsfC{v}{b}{v}{\vk}{\vp}{\vq}&=&\basefunc(\fv_\vk,\fb_\vp,\fv_\vq,\vq)+\circVBV(\fv_\vk,\fb_\vp,\fv_\vq),\\
\label{eq:HcTvvb} \ftsfC{v}{v}{b}{\vk}{\vp}{\vq}&=&-\basefunc(\fv_\vk,\fv_\vp,\fb_\vq,\vq)+\circVVB(\fv_\vk,\fv_\vp,\fb_\vq),
\eeqa

where $(\circBBB,\circBVV,\circVBV,\circVVB)$ are circulating transfer functions.

An equivalent symmetric reformulation of the ideal MHD equations \reqrefssne{eq:mhdS}{eq:mhdE} in terms of the Els\"asser fields $\vzpm=\vv\pm\vb$ is given by \cite{ELS50}:
\beq
\pat \vzpm=-\vzmp\cdot\nabla\vzpm - \nabla p^*, \quad \nabla \cdot \vzpm =0.
\eeq

The energy associated with each Els\"asser field reads $\elsEpm=\frac12|\fzpm|^2=\kinE+\magE\pm\crossH$. The pair transfers of Els\"asser energies are given by (\supeqss{eq:patelsEpm}{eq:zpzpzm}):

\beqa
\label{eq:mtomElsA}\ftsfC{z^{\pm}}{z^{\mp}}{z^{\pm}}{\vk}{\vp}{\vq}&=&-\basefunc(\fzpm_\vk,\fzmp_\vp,\fzpm_\vq,\vq)+\circZpmZmpZpm(\fzpm_\vk,\fzmp_\vp,\fzpm_\vq),\\
\ftsfC{z^{\pm}}{z^{\mp}}{z^{\mp}}{\vk}{\vp}{\vq}&=&\circZpmZmpZmp(\fzpm_\vk,\fzmp_\vp,\fzmp_\vq),\\
\label{eq:mtomElsB} \ftsfC{z^{\pm}}{z^{\pm}}{z^{\mp}}{\vk}{\vp}{\vq}&=&\circZpmZpmZmp(\fzpm_\vk,\fzpm_\vp,\fzmp_\vq),
\eeqa

which involve six circulating transfer terms: $\circZPPM, \circZPMP, \circZPMM, \circZMPP, \circZMPM$ and $\circZMMP$. 

In terms of $(\fzp,\fzm)$, since $\crossH=\frac12(\elsEp-\elsEm)$, the transfer rate of cross helicity, from wave vector $\vk$ to wave vector $\vq$, is given by (\supeqss{eq:wtowEzpkpq}{eq:wtowHckpq}):

\beqa
\nonumber \wtowHc{\vk}{\vp}{\vq}&=&\frac12 \Big(-\basefunc(\fzp_\vk,\fzm_\vp,\fzp_\vq,\vq)+\basefunc(\fzm_\vk,\fzp_\vp,\fzm_\vq,\vq)\\
\nonumber &+&\underbrace{(\circZMPP-\circZPMM)}_{\dcircZk}+\underbrace{(\circZPMP-\circZMPM)}_{\dcircZp}\\
\label{eq:crossHzpm} &-&\underbrace{(\circZPPM-\circZMMP)}_{\dcircZq}\Big).
\eeqa

Equating \reqref{eq:crossHvb} and \reqref{eq:crossHzpm} leads to the following relation between the circulating transfer terms (\supeqss{eq:expandD}{eq:relcircHc}):

\newcommand{\circHcsum}{\circS^\Sigma_{\crossH}}

\beq
\label{eq:circTH} \dcircZk+\dcircZp-\dcircZq=\underbrace{2(\circBBB+\circBVV+\circVBV+\circVVB)}_{\circHcsum}.
\eeq

The same reasoning for the cross helicity transfers from $\vp$ to $\vk$ and from $\vq$ to $\vp$ leads to the system of equations:

\beq
\begin{pmatrix}
1 & 1 & -1 \\
-1 & 1 & 1 \\
1 & -1 & 1
\end{pmatrix}
\begin{pmatrix}
\dcircZk \\
\dcircZp \\
\dcircZq
\end{pmatrix}
=
\begin{pmatrix}
\circHcsum \\
\circHcsum \\
\circHcsum 
\end{pmatrix}
\eeq

which has the unique solution 

\beq
\label{eq:solsys}\dcircZk=\dcircZp=\dcircZq=\circHcsum,
\eeq

giving

\beqa
0&=&\dcircZk-\dcircZp,\\
&=&\circZMPP-\circZPMM-\circZPMP+\circZMPM
\eeqa

As an important consequence of \eqref{eq:R3}, 
any vanishing non-trivial linear combination of transfer functions that all represent different triad interactions, necessarily implies
that each transfer function in this sum has to vanish individually. A rigorous formulation and proof of this result is given in the supplementary material, \reqref{eq:frel}. Hence, since four distinct triads appear ($\{\fzm_\vk,\fzp_\vp,\fzp_\vq\}$ in $\circZMPP$, $\{\fzp_\vk,\fzm_\vp,\fzm_\vq\}$ in $\circZPMM$, $\{\fzp_\vk,\fzm_\vp,\fzp_\vq\}$ in $\circZPMP$ and $\{\fzm_\vk,\fzp_\vp,\fzm_\vq\}$ in $\circZMPM$), and the left-hand side is zero, we deduce that they are all individually identically zero:

\beq
\circZMPP=\circZPMM=\circZPMP=\circZMPM=0.
\eeq

In the same way, $\circZPPM=\circZMMP=0$. Then, \reqref{eq:solsys} implies that $\frac12\circHcsum=\circBBB+\circBVV+\circVBV+\circVVB=0$ and since different triads appear in each circulating transfer function ($\{\fb_\vk,\fb_\vp,\fb_\vq\}, \{\fb_\vk,\fv_\vp,\fv_\vq\}, \{\fv_\vk,\fb_\vp,\fv_\vq\}$ and $\{\fv_\vk,\fv_\vp,\fb_\vq\}$, in this order), all of them are identically zero, as well.

Finally, the transfer of total energy $\totE=\kinE+\magE=\frac12(\elsEp+\elsEm)$ from wave vector $\vk$ to $\vq$ in terms of $(\fv,\fb)$ is given by

\beq
\label{eq:crossEtvb} \wtowEt{\vk}{\vp}{\vq}=\Big(\ftsfform{v}{v}{v}+\ftsfform{v}{b}{b}+\ftsfform{b}{v}{b}+\ftsfform{b}{b}{v}\Big)(\ftsfCarg{\vk}{\vp}{\vq}),
\eeq

In terms of $(\fzp,\fzm)$, since the $\circZaZbZc$ are zero, this is equivalent to

\beq
\label{eq:crossEtzpm} \wtowEt{\vk}{\vp}{\vq}=\frac12\Big(-\basefunc(\fzp_\vk,\fzm_\vp,\fzp_\vq,\vq)-\basefunc(\fzm_\vk,\fzp_\vp,\fzm_\vq,\vq)\Big).
\eeq

Equating \reqref{eq:crossEtvb} and \reqref{eq:crossEtzpm} yields (\supeqss{eq:wtowEtkpq}{eq:relcircEt}):

\beq
\label{eq:circTE}\circVVV+\circVBB+\circBVB+\circBBV=0
\eeq

and since each circulating transfer function is a function of different triads ($\{\fv_\vk,\fv_\vp,\fv_\vq\}, \{\fv_\vk,\fb_\vp,\fb_\vq\}, \{\fb_\vk,\fv_\vp,\fb_\vq\}$ and $\{\fb_\vk,\fb_\vp,\fv_\vq\}$, in this order), each is identically zero.

Thus, all circulating transfer functions are identically zero. The only possible pair transfer functions are given by \reqrefss{eq:generalvv}{eq:generalbb}, without the circulating transfer functions on the right-hand side.

\section{Conclusion}

The uniqueness of pair transfer functions in incompressible MHD justifies and motivates their use in theoretical and experimental studies. Thus, the locality, direction and nature of energetic transfers is mathematically unambiguous for triply periodic systems.

The uniqueness of the kinetic$\lra$kinetic energetic exchanges in MHD leads to a natural question about the hydrodynamic case $|\vb|=0$. The pair transfer functions are continuous in all of their arguments. Hence, the only possible prolongation by continuity in the hydrodynamic case is the one given by \reqref{eq:generalvv} with $\circVVV=0$. The question remains whether such a passage to the limit $|\vb|\to 0$ would be justified, since the proof presented here relies on the transfers of cross helicity and the distinguishability of both Els\"asser fields $\vv\pm\vb$.

\begin{acknowledgments}
Fruitful discussions with Raquel M\"ausle and Angela Busse are thankfully acknowledged.
\end{acknowledgments}


\bibliography{biblio.bib}

@Book{leslie:book,
  author    = {D. C. Leslie},
  publisher = {Clarendon Press},
  title     = {Developments in the Theory of Turbulence},
  year      = {1983},
  address   = {Oxford},
}

@InCollection{orszag:stattheory,
  author    = {S. A. Orszag},
  booktitle = {Fluid Dynamics/Dynamique des Fluides},
  publisher = {Gordon and Breach Science Publishers},
  title     = {Lectures on the Statistical Theory of Turbulence},
  year      = {1973},
  editor    = {R. Balian and J.-L. Peube},
  pages     = {236--374},
  series    = {Cours de l'Ecole d'Et\'e de Physique Th\'eorique, Les Houches},
}

@Book{monin_yaglom:book2,
  author    = {A. S. Monin and A. M. Yaglom},
  publisher = {{MIT} Press},
  title     = {Statistical Fluid Mechanics},
  year      = {1981},
  address   = {Cambridge, Massachusetts},
  volume    = {2}
}

@Article{kraichnan:dia,
  author  = {R. H. Kraichnan},
  journal = {Journal of Fluid Mechanics},
  title   = {The structure of isotropic turbulence at very high {R}eynolds numbers},
  year    = {1959},
  pages   = {497--543},
  volume  = {5}
}

@Article{kraichnan:lhdia,
  author  = {R. H. Kraichnan},
  journal = {Physics of Fluids},
  title   = {Lagrangian-history closure approximation for turbulence},
  year    = {1965},
  number  = {4},
  pages   = {575--598},
  volume  = {8}
}

@Article{beck_mueller.networkmodel,
  author    = {Benjamin Beck and Wolf-Christian Müller},
  journal   = {Journal of Physics: Conference Series},
  title     = {Network model of magnetohydrodynamic turbulence},
  year      = {2025},
  issn      = {1742-6596},
  number    = {1},
  pages     = {012002},
  volume    = {2997},
  doi       = {10.1088/1742-6596/2997/1/012002},
  publisher = {IOP Publishing}
}

@Article{alexakis_mininni_pouquet:eflowdiag,
  author  = {A. Alexakis and P. D. Mininni and A. Pouquet},
  journal = {New Journal of Physics},
  title   = {Turbulent cascades, transfer and scale-interactions in magnetohydrodynamics},
  year    = {2007},
  pages   = {298},
  volume  = {9}
}

@Article{verma:review,
  author  = {M. K. Verma},
  journal = {Physics Reports},
  title   = {Statistical theory of magnetohydrodynamic turbulence: recent results},
  year    = {2004},
  pages   = {229--380},
  volume  = {401},
}

@article{DVE01,
   author = {Gaurav Dar and Mahendra K. Verma and V. Eswaran},
   doi = {10.1016/S0167-2789(01)00307-4},
   issn = {01672789},
   issue = {3},
   journal = {Physica D: Nonlinear Phenomena},
   month = {9},
   pages = {207-225},
   publisher = {North-Holland},
   title = {Energy transfer in two-dimensional magnetohydrodynamic turbulence: formalism and numerical results},
   volume = {157},
   url = {https://www.sciencedirect.com/science/article/pii/S0167278901003074},
   year = {2001}
}

@article{ELS50,
   author = {Walter M. Elsasser},
   doi = {10.1103/PhysRev.79.183},
   issn = {0031899X},
   issue = {1},
   journal = {Physical Review},
   month = {7},
   pages = {183},
   publisher = {American Physical Society},
   title = {The Hydromagnetic Equations},
   volume = {79},
   url = {https://journals.aps.org/pr/abstract/10.1103/PhysRev.79.183},
   year = {1950}
}

@article{PSV19,
   author = {Franck Plunian and Rodion Stepanov and Mahendra Kumar Verma},
   doi = {10.1017/S0022377819000710},
   issn = {0022-3778},
   issue = {5},
   journal = {Journal of Plasma Physics},
   month = {10},
   pages = {905850507},
   publisher = {Cambridge University Press},
   title = {On uniqueness of transfer rates in magnetohydrodynamic turbulence},
   volume = {85},
   url = {https://www.cambridge.org/core/journals/journal-of-plasma-physics/article/abs/on-uniqueness-of-transfer-rates-in-magnetohydrodynamic-turbulence/F50056240BD2F9A1AF0A1CB1A4C0AB2F},
   year = {2019}
}

\clearpage

\appendix
\section{Supplementary material}
This supplement first reviews the effect of nonlinear terms of the form $(\vg\cdot\nabla) \vf$ on energetic transfers and
discusses the conservative transfer between modes in a triad interaction in incompressible MHD.

Considering in addition the transfers of cross helicity and of Els\"asser energies leads to general expressions
for the pair transfers, including the circulating terms.

Finally, equating the transfers of cross helicity and of total energy in both equivalent views (i.e. in terms of $(\fv,\fb)$ and in terms of $(\fzp,\fzm)$) yields relations between the circulating transfer functions. These relations can hold if and only if the circulating transfer functions are identically zero, as shown in the main text.

The proof of this statement requires a specific  property of the nonlinear transfer functions of incompressible MHD: any vanishing linear combination of transfer functions that all represent different triad interactions, necessarily implies
that each transfer function in this sum has to vanish individually, provided they fulfill requirement R3.
This central result is formalized and proven at the end of this supplementary material. To show that requirement R3 is not a consequence of R1 and R2, we provide lastly counter-examples of such circulating transfer functions.

\subsection{Nonlinear rate of change of the energy}

Consider the nonlinear evolution equation
\[\pat \vh = \pm (\vg \cdot \nabla)\vf\] in a finite triply-periodic volume.

It has the corresponding Fourier representation
\[\pat \fh_\vq = \pm i\sum_{\vk+\vp=\vq} ( \fg_\vp \cdot \vk)\ff_\vk,\]
with e.g. $\ff_\vk$ the Fourier amplitude at wave vector $\vk$. 

With $\somEh_\vq=\frac12 |\fh_\vq|^2=\frac12 \fh_\vq \cdot \fh^*_\vq$, the asterisk $^*$ denoting the complex conjugate, the energy evolution is governed by

\beqa
\pat \somEh_\vq&=&\frac12 \Big((\pat \fh_\vq) \cdot \fh^*_\vq+ \fh_\vq \cdot \pat \fh^*_\vq \Big),\label{s0}\\
&=&\mp \Im\Big( \sum_{\vk+\vp=\vq} ( \fg_\vp \cdot \vk) (\ff_\vk \cdot \fh^*_\vq) \Big),\\
&=&\mp \sum_{\vk+\vp+\vq=0}  \Im\Big( ( \vk \cdot \fg_\vp ) (\ff_\vk \cdot \fh_\vq) \Big),\label{s1}\\
&=& \pm \sum_{\vk+\vp+\vq=0}  \Im\Big( ( \vq \cdot \fg_\vp ) (\ff_\vk \cdot \fh_\vq) \Big)\label{eq:nonsympatEhvq},\\
\label{eq:finalpatEhvq}&=& \pm \frac12 \sum_{\vk+\vp+\vq=0}  \Big( \basefunc(\ff_\vk,\fg_\vp,\fh_\vq,\vq)+\basefunc(\ff_\vp,\fg_\vk,\fh_\vq,\vq) \Big).
\eeqa

Here, Eq. \eqref{s1} is due to Hermitian symmetry and Eq. \eqref{eq:nonsympatEhvq} to the solenoidality of the field $\fg$, i.e. $\vp\cdot\fg_\vp=0$.
The final form, Eq. \eqref{eq:finalpatEhvq}, exploits the symmetry in $\vk$ and $\vp$
with $\basefunc(\ff_\vk,\fg_\vp,\fh_\vq,\vq)=\Im(\tsfIMterm{\ff}{\fg}{\fh}{\vk}{\vp}{\vq})$.

\subsection{Conservative energy transfer by triad interactions (\reqref{eq:detailedvvv} and \reqref{eq:detailedvbb})}

For the $\{\fv_\vk,\fv_\vp,\fv_\vq\}$ triad, one has $\fhypbudget{\fv_\vk}{\fv_\vp}{\fv_\vq}=\fhypvgv{\vk}{\vp}{\vq}$. Thus, using \reqref{eq:hypervgv}:

\newcommand{\dotbox}[1]{%
  \tikz[baseline=(char.base)]
    \node[anchor=base, draw=black, dotted, inner sep=2pt, rectangle] (char) {#1};%
}

\newcommand{\dashedbox}[1]{%
  \tikz[baseline=(char.base)]
    \node[anchor=base, draw=black, dashed, inner sep=2pt, rectangle] (char) {#1};%
}

\newcommand{\solidbox}[1]{%
  \tikz[baseline=(char.base)]
    \node[anchor=base, draw=black, inner sep=2pt, rectangle] (char) {#1};%
}
 
\newcommand{\boxtA}[1]{\solidbox{$\displaystyle #1$}}
\newcommand{\boxtB}[1]{\dashedbox{$\displaystyle #1$}}
\newcommand{\boxtC}[1]{\dotbox{$\displaystyle #1$}}

\beqa
\label{eq:supmat_vkvpvq}\fhypbudget{\fv_\vk}{\fv_\vp}{\fv_\vq}&=&\boxtA{-\Im(\tsfIMterm{\fv}{\fv}{\fv}{\vk}{\vp}{\vq})}\,\boxtC{-\Im(\tsfIMterm{\fv}{\fv}{\fv}{\vp}{\vk}{\vq}))},\\
\fhypbudget{\fv_\vp}{\fv_\vq}{\fv_\vk}&=&\boxtB{-\Im(\tsfIMterm{\fv}{\fv}{\fv}{\vp}{\vq}{\vk})}\,\boxtA{-\Im(\tsfIMterm{\fv}{\fv}{\fv}{\vq}{\vp}{\vk}))},\\
\fhypbudget{\fv_\vq}{\fv_\vk}{\fv_\vp}&=&\boxtC{-\Im(\tsfIMterm{\fv}{\fv}{\fv}{\vq}{\vk}{\vp})}\,\boxtB{-\Im(\tsfIMterm{\fv}{\fv}{\fv}{\vk}{\vq}{\vp}))}.
\eeqa

In the sum $\fhypbudget{\fv_\vk}{\fv_\vp}{\fv_\vq}+\fhypbudget{\fv_\vp}{\fv_\vq}{\fv_\vk}+\fhypbudget{\fv_\vq}{\fv_\vk}{\fv_\vp}$, the terms in the same box style cancel pairwise when using $\kpqz$ and $\vm \cdot\fv_\vm=0$.

Regarding triad interactions involving a velocity mode and two magnetic ones, e.g. $\{\fb_\vk,\fb_\vp,\fv_\vq\}$, the $\fhypbudget{\fb_\vk}{\fb_\vp}{\fv_\vq}$ budget corresponds to $\fhypbgb{\vk}{\vp}{\vq}$, hence:

\beq
\fhypbudget{\fb_\vk}{\fb_\vp}{\fv_\vq}=\boxtB{\Im(\tsfIMterm{\fb}{\fb}{\fv}{\vk}{\vp}{\vq})}+\boxtC{\Im(\tsfIMterm{\fb}{\fb}{\fv}{\vp}{\vk}{\vq})}.
\eeq

Contrary to the budgets above, $\fhypbudget{\fb_\vp}{\fv_\vq}{\fb_\vk}$ and $\fhypbudget{\fv_\vq}{\fb_\vk}{\fb_\vp}$ do not correspond to a single $\fhypggg$ term but to both terms in the induction equation (cf \reqrefss{eq:hypervgv}{eq:hyperbgv}):

\beqa
\fhypbudget{\fb_\vp}{\fv_\vq}{\fb_\vk}=\boxtA{-\Im(\tsfIMterm{\fb}{\fv}{\fb}{\vp}{\vq}{\vk})}+\boxtB{\Im(\tsfIMterm{\fv}{\fb}{\fb}{\vq}{\vp}{\vk}))},\\
\label{eq:supmat_vqbkbp}\fhypbudget{\fv_\vq}{\fb_\vk}{\fb_\vp}=\boxtA{-\Im(\tsfIMterm{\fb}{\fv}{\fb}{\vk}{\vq}{\vp})}+\boxtC{\Im(\tsfIMterm{\fv}{\fb}{\fb}{\vq}{\vk}{\vp}))}.
\eeqa

Similarly to the case of the $\{\fv_\vk,\fv_\vp,\fv_\vq\}$ triad, the terms in the sum $\fhypbudget{\fb_\vk}{\fb_\vp}{\fv_\vq}+\fhypbudget{\fb_\vp}{\fv_\vq}{\fb_\vk}+\fhypbudget{\fv_\vq}{\fb_\vk}{\fb_\vq}$ cancel pairwise, as indicated by the box styles.

\subsection{Pair transfers of cross helicity (\reqrefss{eq:crossHvb}{eq:HcTvvb})}

The cross helicity at wave vector $\vq$ is governed by:

\newcommand{\cc}{\mathit{c.c.}}

\beqa
\label{eq:patcrossH}\pat \crossH_\vq&=&\frac12 \Big((\pat \fv_\vq) \cdot \fb^*_\vq + (\pat \fb_\vq) \cdot \fv^*_\vq\Big) + \cc,\\
\label{eq:patcrossHRe}&=&\Re\Big((\pat \fv_\vq) \cdot \fb^*_\vq + (\pat \fb_\vq) \cdot \fv^*_\vq\Big),
\eeqa

`$\cc$' denoting the complex conjugate of the terms to its left.
Transforming \reqref{eq:patcrossHRe} in a manner similar to the manipulations following Eq. \eqref{s0} gives

\beqa
\nonumber \pat \crossH_\vq&=&\pm \frac12 \sum_{\vk+\vp+\vq=0}  \Big( \fhypHcvgv{\vk}{\vp}{\vq}+\fhypHcbgb{\vk}{\vp}{\vq} \\
\label{eq:finaldtHc} &+&\fhypHcvgb{\vk}{\vp}{\vq}+\fhypHcbgv{\vk}{\vp}{\vq}\Big),
\eeqa

with

\beqa
\label{eq:fhypHcvgv}\fhypHcvgv{\vk}{\vp}{\vq}&=&-\basefunc(\fv_\vk,\fv_\vp,\fb_\vq,\vq)-\basefunc(\fv_\vp,\fv_\vk,\fb_\vq,\vq),\\
\fhypHcbgb{\vk}{\vp}{\vq}&=&\basefunc(\fb_\vk,\fb_\vp,\fb_\vq,\vq)+\basefunc(\fb_\vp,\fb_\vk,\fb_\vq,\vq),\\
\fhypHcvgb{\vk}{\vp}{\vq}&=&-\basefunc(\fb_\vk,\fv_\vp,\fv_\vq,\vq)-\basefunc(\fb_\vp,\fv_\vk,\fv_\vq,\vq),\\
\label{eq:fhypHcbgb}\fhypHcbgb{\vk}{\vp}{\vq}&=&\basefunc(\fv_\vk,\fb_\vp,\fv_\vq,\vq)+\basefunc(\fv_\vp,\fb_\vk,\fv_\vq,\vq),
\eeqa

Thus, in a triangle $\kpqz$, four triads are responsible for the transfers of cross helicity: $\{\fb_\vk,\fb_\vp,\fb_\vq\}, \{\fb_\vk,\fv_\vp,\fv_\vq\}, \{\fv_\vk,\fb_\vp,\fv_\vq\}$ and $\{\fv_\vk,\fv_\vp,\fb_\vq\}$.

The rate of change of cross helicity at wave vector $\vq$ due to the triad of three magnetic modes is:

\beq
\fhypbudget{\fb_\vk}{\fb_\vp}{\fb_\vq}=\basefunc(\fb_\vk,\fb_\vp,\fb_\vq,\vq)+\basefunc(\fb_\vp,\fb_\vk,\fb_\vq,\vq).
\eeq

The exact same derivations as in the case of kinetic energy transfers among the $\{\fv_\vk,\fv_\vp,\fv_\vq\}$ triad allow to conclude that this interaction conservatively redistributes the cross helicity among the wave vectors, i.e. $\fhypbudget{\fb_\vk}{\fb_\vp}{\fb_\vq}+\fhypbudget{\fb_\vp}{\fb_\vq}{\fb_\vk}+\fhypbudget{\fb_\vq}{\fb_\vk}{\fb_\vp}=0$. This allows to define in all generality the corresponding pair transfer function:

\beq
\ftsfC{b}{b}{b}{\vk}{\vp}{\vq}=\basefunc(\fb_\vk,\fb_\vp,\fb_\vq,\vq)+\circBBB(\fb_\vk,\fb_\vp,\fb_\vq),
\eeq

with $\circBBB$ a circulating transfer function.

For a triad of two velocity modes and one magnetic one, e.g. $\{\fv_\vk,\fv_\vp,\fb_\vq\}$, \reqrefss{eq:finaldtHc}{eq:fhypHcbgb} result in: 

\beqa
\fhypbudget{\fv_\vk}{\fv_\vp}{\fb_\vq}&=&\boxtA{-\basefunc(\fv_\vk,\fv_\vp,\fb_\vq,\vq)}\,\boxtB{-\basefunc(\fv_\vp,\fv_\vk,\fb_\vq,\vq)},\\
\fhypbudget{\fv_\vp}{\fb_\vq}{\fv_\vk}&=&\boxtA{-\basefunc(\fb_\vq,\fv_\vp,\fv_\vk,\vk)}+\boxtC{\basefunc(\fv_\vp,\fb_\vq,\fv_\vk,\vk)},\\
\fhypbudget{\fb_\vq}{\fv_\vk}{\fv_\vp}&=&\boxtB{-\basefunc(\fb_\vq,\fv_\vk,\fv_\vp,\vp)}+\boxtC{\basefunc(\fv_\vk,\fb_\vq,\fv_\vp,\vp)}.
\eeqa

The same derivations as for triads of two magnetic modes and one velocity one in energetic transfers allows to conclude that in the sum $\fhypbudget{\fv_\vk}{\fv_\vp}{\fb_\vq}+\fhypbudget{\fv_\vp}{\fb_\vq}{\fv_\vk}+\fhypbudget{\fb_\vq}{\fv_\vk}{\fv_\vp}$ the terms cancel pairwise, as indicated by the box styles. One can thus define for the triad $\{\fv_\vk,\fv_\vp,\fb_\vq\}$ in all generality the following pair cross helicity transfer functions:

\beqa
\ftsfC{b}{v}{v}{\vq}{\vk}{\vp}&=&-\basefunc(\fb_\vq,\fv_\vk,\fv_\vp,\vp)+\circVVB(\fv_\vk,\fv_\vp,\fb_\vq),\\
\ftsfC{v}{b}{v}{\vp}{\vq}{\vk}&=&\basefunc(\fv_\vp,\fb_\vq,\fv_\vk,\vk)+\circVVB(\fv_\vk,\fv_\vp,\fb_\vq),\\
\label{eq:vvbwtot} \ftsfC{v}{v}{b}{\vk}{\vp}{\vq}&=&-\basefunc(\fv_\vk,\fv_\vp,\fb_\vq,\vq)+\circVVB(\fv_\vk,\fv_\vp,\fb_\vq),
\eeqa

with $\circVVB$ the corresponding circulating transfer function. The same reasoning for the remaining $\{\fv_\vk,\fb_\vp,\fv_\vq\}$ and $\{\fb_\vk,\fv_\vp,\fv_\vq\}$ triads gives the same form of pair cross helicity transfer functions, but with the other circulating transfer functions $\circVBV$ and $\circBVV$.

\subsection{Pair transfers of Els\"asser energies (\reqrefss{eq:mtomElsA}{eq:mtomElsB})}

Using \reqref{eq:nonsympatEhvq}, the Els\"asser energies in a single triangle $\kpqz$ are governed by:

\beq
\label{eq:patelsEpm} \pat \elsEpm_\vq = -\Im\Big( ( \vq \cdot \fzmp_\vp ) (\fzpm_\vk \cdot \fzpm_\vq) \Big)-\Im\Big( ( \vq \cdot \fzmp_\vk ) (\fzpm_\vp \cdot \fzpm_\vq) \Big).
\eeq

Thus, only triads involving both Els\"asser signs play a role. Thus, there are six of them in a triangle: $\{\fzm_\vk,\fzp_\vp,\fzp_\vq\}, \{\fzp_\vk,\fzm_\vp,\fzp_\vq\}, \{\fzp_\vk,\fzp_\vp,\fzm_\vq\}, \{\fzp_\vk,\fzm_\vp,\fzm_\vq\},\{\fzm_\vk,\fzp_\vp,\fzm_\vq\}$ and $\{\fzm_\vk,\fzm_\vp,\fzp_\vq\}$. Since they all have the same form: two Els\"asser fields of one Els\"asser sign and one of the opposite sign, it is sufficient to consider only one such triad, say $\{\fzp_\vk,\fzm_\vp,\fzp_\vq\}$.

When considering this triad interaction in isolation, \reqref{eq:patelsEpm} shows that the energetic content of $\fzm_\vp$ does not change. A change of $\elsEm_\vp$ would indeed require a second mode with a negative Els\"asser sign. We read from \reqref{eq:patelsEpm}:

\beqa
\fhypbudget{\fzp_\vk}{\fzm_\vp}{\fzp_\vq}&=&-\Im\Big( ( \vq \cdot \fzm_\vp ) (\fzp_\vk \cdot \fzp_\vq) \Big),\\
\fhypbudget{\fzm_\vp}{\fzp_\vq}{\fzp_\vk}&=&-\Im\Big( ( \vk \cdot \fzm_\vp ) (\fzp_\vq \cdot \fzp_\vk) \Big),\\
\fhypbudget{\fzp_\vq}{\fzp_\vk}{\fzm_\vp}&=&0.
\eeqa

We have $\fhypbudget{\fzp_\vk}{\fzm_\vp}{\fzp_\vq}=-\fhypbudget{\fzm_\vp}{\fzp_\vq}{\fzp_\vk}$ and thus this triad interaction conserves the Els\"asser energies. Thus, the pair transfer functions are in all generality:

\beqa
\label{t1}\ftsfC{z^+}{z^-}{z^+}{\vk}{\vp}{\vq}&=&-\basefunc(\fzp_\vk,\fzm_\vp,\fzp_\vq,\vq)+\circZPMP(\fzp_\vk,\fzm_\vp,\fzp_\vq),\\
\ftsfC{z^+}{z^+}{z^-}{\vq}{\vk}{\vp}&=&\circZPMP(\fzp_\vk,\fzm_\vp,\fzp_\vq),\\
\label{eq:zpzpzm}\ftsfC{z^-}{z^+}{z^+}{\vp}{\vq}{\vk}&=&\circZPMP(\fzp_\vk,\fzm_\vp,\fzp_\vq).\label{t3}
\eeqa

The pair transfer functions for the five remaining triads are obtained by interchanging the Els\"asser signs and by circular permutation of the wave vectors. Each involves an \textit{a priori} different circulating transfer function.

\subsection{Relations between circulating terms (\reqref{eq:crossHzpm}, \reqref{eq:circTH} and \reqref{eq:circTE})}

From the previous section, the pair transfer of Els\"asser energies follows as:

\beqa
\label{eq:wtowEzpkpq}\wtowEzp{\vk}{\vp}{\vq}&=&\Big(\ftsfform{z^+}{z^-}{z^+}+\ftsfform{z^-}{z^+}{z^+}+\ftsfform{z^-}{z^-}{z^+}\Big)(\ftsfCarg{\vk}{\vp}{\vq}),\\
\wtowEzm{\vk}{\vp}{\vq}&=&\Big(\ftsfform{z^-}{z^+}{z^-}+\ftsfform{z^+}{z^-}{z^-}+\ftsfform{z^+}{z^+}{z^-}\Big)(\ftsfCarg{\vk}{\vp}{\vq}).
\eeqa

Since the cross helicity can also be expressed as $\crossH_\vq=\frac12(\elsEp_\vq-\elsEm_\vq)$:

\beq
\label{eq:patcrossHdefEz}
\wtowHc{\vk}{\vp}{\vq}=\frac12\Big(\wtowEzp{\vk}{\vp}{\vq}-\wtowEzm{\vk}{\vp}{\vq}\Big).
\eeq

Using Eqs. \eqref{t1}--\eqref{t3} in \reqref{eq:patcrossHdefEz} and grouping the circulating terms two by two gives \reqref{eq:crossHzpm}: 

\beqa
\nonumber \wtowHc{\vk}{\vp}{\vq}&=&\frac12 \Big(-\basefunc(\fzp_\vk,\fzm_\vp,\fzp_\vq,\vq)+\basefunc(\fzm_\vk,\fzp_\vp,\fzm_\vq,\vq)\\
\label{eq:wtowHckpq}&+&\dcircZk+\dcircZp-\dcircZq\Big).
\eeqa

A partial reformulation of \eqref{eq:wtowHckpq} in terms of the $(\fv,\fb)$ fields yields a relation between the circulating transfer terms from both perspectives.
Since $\basefunc$ is linear in all of its arguments, this results in:

\beqa
\nonumber\basefunc(\fzpm_\vk,\fzmp_\vp,\fzpm_\vq,\vq)&=&\basefunc(\fv_\vk,\fv_\vp,\fv_\vq,\vq)-\basefunc(\fv_\vk,\fb_\vp,\fb_\vq,\vq)\\
\nonumber&+&\basefunc(\fb_\vk,\fv_\vp,\fb_\vq,\vq)-\basefunc(\fb_\vk,\fb_\vp,\fv_\vq,\vq)\\
\nonumber&\pm&\basefunc(\fv_\vk,\fv_\vp,\fb_\vq,\vq)\mp\basefunc(\fv_\vk,\fb_\vp,\fv_\vq,\vq)\\
\label{eq:expandD}&\pm&\basefunc(\fb_\vk,\fv_\vp,\fv_\vq,\vq)\mp\basefunc(\fb_\vk,\fb_\vp,\fb_\vq,\vq).
\eeqa

Thus, when considering the sum $\basefunc(\fzp_\vk,\fzm_\vp,\fzp_\vq,\vq)+\basefunc(\fzm_\vk,\fzp_\vp,\fzm_\vq,\vq)$, only the triads relevant for energetic transfers (with three velocity modes or one velocity and two magnetic ones) remain and, conversely, when considering the difference, only those relevant for cross helicity transfers (with three magnetic modes or one magnetic and two velocity ones) do remain.

Thus, when equating the transfer rate of cross helicity given by \reqref{eq:crossHvb} and \reqref{eq:crossHzpm}, the terms in $\basefunc$ cancel out and we are left with a relation between the circulating transfer terms from both, the $(\fv,\fb)$ and $(\fzp,\fzm)$ perspectives:

\beq
\label{eq:relcircHc}\dcircZk+\dcircZp-\dcircZq=2(\circBBB+\circBVV+\circVBV+\circVVB).
\eeq

Similarly, the pair transfer of total energy is given by:

\beq
\label{eq:wtowEtkpq}\wtowEt{\vk}{\vp}{\vq}=\frac12\Big(\wtowEzp{\vk}{\vp}{\vq}+\wtowEzm{\vk}{\vp}{\vq}\Big).
\eeq

When equating it to \reqref{eq:crossEtvb}, the terms in $\basefunc$ cancel out and we are left with:

\beqa
\nonumber 2(\circVVV+\circVBB+\circBVB+\circBBV)&=&\circZPPM+\circZPMP+\circZPMM\\
\label{eq:relcircEt}&+&\circZMMP+\circZMPM+\circZMPP.
\eeqa

\subsection{Independence of transfer functions associated with different triads}
\newcommand{\fza}{\hat{\bf z}}
\newcommand{\fzI}{\fza_1}
\newcommand{\fzII}{\fza_2}
\newcommand{\fzIII}{\fza_3}
\newcommand{\fzIV}{\fza_4}
\newcommand{\fzV}{\fza_5}
\newcommand{\fzVI}{\fza_6}

\newcommand{\fxa}{{\bf x}}
\newcommand{\fxaj}{\fxa_j}
\newcommand{\fxI}{\fxa_1}
\newcommand{\fxII}{\fxa_2}
\newcommand{\fxIII}{\fxa_3}
\newcommand{\fxIV}{\fxa_4}
\newcommand{\fxV}{\fxa_5}
\newcommand{\fxVI}{\fxa_6}
\newcommand{\fxVII}{\fxa_7}
\newcommand{\fxVIII}{\fxa_8}

\newcommand{\tf}{\tilde{f}}

\newcommand{\trifunc}{\sigma}
\newcommand{\absmode}[1]{{\mathcal M}_{#1}} 
\newcommand{\absmodez}[1]{{\mathcal Z}_{#1}}

\newcommand{\intinterv}[2]{\llbracket #1,#2 \rrbracket}
In the present work, it is essential that any finite set formed by transfer functions of $N>1$ different triads,

$\mathbbm{T}_N=\{\ftsfC{\alpha_{j}}{\beta_{j}}{\gamma_{j}}{\vk}{\vp}{\vq}$, $1\leq j\leq N$\},
is independent. A triad is denoted symbolically by
a tuple of field identificators

($\alpha_{j}$,$\beta_{j}$,$\gamma_{j}$) with $\alpha_j,\beta_j,\gamma_j\in\{v,b\} \text{ or alternatively} \in\{z^+,z^-\}$
where $\alpha$ is the mode at wave vector $\vk$, $\beta$ at $\vp$ and $\gamma$ at $\vq$.

For the sake of notational simplicity, the explicit dependency of $\ftsf$ on the wave vectors is dropped.  Setting any of the
field identificator symbols to zero indicates a vanishing amplitude of the corresponding component.  

Independence of $\mathbbm{T}_N$ implies that the equation
\beqa
\sum_{1\leq j\leq N} \ftsfform{\alpha_{j}}{\beta_{j}}{\gamma_{j}}=0 \label{eq:frel}
\eeqa
only holds, if each summand vanishes individually. 

\paragraph{Proof}
In three-dimensional incompressible MHD for a fixed triangle of wave vectors $\vk$, $\vp$, $\vq$, consider a set of different triads, $\mathbbm{T}_{N>1}$,
i.e. for any two different members of the set
the  triad identificators are not identical.
In addition, the corresponding transfer functions fulfill  
the property \eqref{eq:R3}, i.e. $\ftsfform{(\alpha_j=0)}{\beta_j}{\gamma_j}=\ftsfform{\alpha_j}{(\beta_j=0)}{\gamma_j}=
\ftsfform{\alpha_j}{\beta_j}{(\gamma_j=0)}=0$. Eq. \eqref{eq:frel} is supposed to be valid for any physically possible set of mode amplitudes,
in particular for vanishing mode amplitudes.
Thus, given any specific member of the set, say with $j=j*$, all other members  involve at least one field  that is not present at the same wave vector,
as in the triad with $j=j*$. Setting those modes to zero for all $j\neq j^*$ implies for Eq. \eqref{eq:frel}, due to \eqref{eq:R3}, that       
$\ftsfform{\alpha_{j*}}{\beta_{j*}}{\gamma_{j*}}=0$. Consequently, all members of $\mathbbm{T}_N$ vanish individually.

\subsection{About requirement R3}

Multilinearity in the three interacting modes $\ff_\vk,\fg_\vp,\fh_\vq$ is a sufficient condition for requirement R3. However, although the $\genfhypbudget$ functions are multilinear in the three modes, there is no such necessity for the circulating transfer functions leading to solutions of the system of equations \eqref{eq:sysA}. We provide here some counter-examples which are compatible with requirements R1 and R2 for pair transfers but do not fulfill requirement R3.

Consider:

\newcommand{\LTn}[1]{| #1 |}
\newcommand{\sLTn}[1]{\LTn{#1}^{\frac12}}
\newcommand{\tLTn}[1]{\LTn{#1}^{\frac13}}

\beqa
\nonumber \circf{f}{g}{h}&=&(\tLTn{\vk}\LTn{\ff_\vk}-\tLTn{\vp}\LTn{\fg_\vp})(\tLTn{\vp}\LTn{\fg_\vp}-\tLTn{\vq}\LTn{\fh_\vq})\\
\label{eq:RTa}&\times&(\tLTn{\vq}\LTn{\fh_\vq}-\tLTn{\vk}\LTn{\ff_\vk}),
\eeqa

with $\LTn{\fz}=\sqrt{\fz\cdot\fz^*}$. For notational simplicity, the arguments of the function, $(\ff_\vk,\fg_\vp,\fh_\vq,\vk,\vp,\vq)$, have been omitted. This circulating transfer function has the right dimension for an energy transfer rate, fulfills R1 and R2, but is not zero when one of the amplitudes is zero.

If one does not require an explicit dependence on the mode amplitudes, the following transfer function represents a second possible counter-example:

\beq
\label{eq:RTb} \circf{f}{g}{h}\sim\vm \cdot (\vk \times \vq),
\eeq

with $\vm$ any vector that is not in the plane spanned by $\vk,\vp,\vq$.

\end{document}